\documentclass[twocolumn,tighten,times]{aastex631}

\usepackage{amsmath}
\usepackage{amssymb}
\usepackage[T1]{fontenc}

\newcommand\cts{counts~s$^{-1}$}

\shorttitle{Bayesian X-ray Polarimetry Reveals Fast Polarization Angle Variations}
\shortauthors{Li et al.}

\graphicspath{{./}{fig/}}

\begin{document}

\title{X-ray Polarimetry in the Low Statistics Regime using the Bayesian Approach Reveals Polarization Angle Variations}

\author[0000-0002-7340-2188]{Hong Li}
\affiliation{State Key Laboratory of Particle Astrophysics, Institute of High Energy Physics, Chinese Academy of Sciences, Beijing 100049, China} 

\author[0000-0001-9893-8248]{Qing-Chang Zhao}
\affiliation{State Key Laboratory of Particle Astrophysics, Institute of High Energy Physics, Chinese Academy of Sciences, Beijing 100049, China} 
\affiliation{University of Chinese Academy of Sciences, Chinese Academy of Sciences, Beijing 100049, China} 

\author[0000-0001-7584-6236]{Hua Feng}
\email{hfeng@ihep.ac.cn}
\affiliation{State Key Laboratory of Particle Astrophysics, Institute of High Energy Physics, Chinese Academy of Sciences, Beijing 100049, China} 

\author[0000-0002-2705-4338]{Lian Tao}
\email{taolian@ihep.ac.cn}
\affiliation{State Key Laboratory of Particle Astrophysics, Institute of High Energy Physics, Chinese Academy of Sciences, Beijing 100049, China} 

\author[0000-0002-9679-0793]{Sergey S. Tsygankov}
\affiliation{Department of Physics and Astronomy, University of Turku, 20014 Turku, Finland}

\begin{abstract}
X-ray polarimetry of accreting compact object has revealed fast time variations in the polarization angle (PA), suggesting that the geometry and/or optical depth of the Comptonization region is changing rapidly.
This prompts investigations into how fast such variability can be.
Conventionally, the data are often binned to examine the time variability such that the measurement in each bin is above the minimum detectable polarization (MDP).
Here we demonstrate that this is unnecessary, and even below the MDP, one can infer the posterior distribution of PA reliably using the Bayesian approach and still be able to place useful constraints on the physics in many cases, due to small relative uncertainties on PA (e.g., $\Delta {\rm PA} \approx 10-30\arcdeg$ compared with a dynamical range of 180\arcdeg). 
With this approach, we discovered that the PA variation in one of the Imaging X-ray Polarimetry Explorer (IXPE) observations of GX 13+1 is not following a linear rotation mode as suggested previously. 
Instead, the PA swings between two discrete angles, suggesting that there are two emitting components, e.g., the boundary layer and the spreading layer, competing with each other.
In XTE J1701$-$462, we confirmed previous results for a variable PA in the normal branch, and furthermore, revealed that the variation timescale could be as short as 1.5 hours. 
During the IXPE observation of Sco X-1, a hint is found for the PA in the highest flux level to be different from the average but consistent with previous measurement results with PolarLight and OSO-8. 
\end{abstract}

\section{Introduction}

X-ray polarimetry serves as a diagnostic tool for constraining the emission mechanism, magnetic field and radiative transfer geometry in high energy astrophysics \citep{kallman2004, kim2024,taverna2024,soffitta2024,IXPE_BH_Review,IXPE_PWN_Review,IXPE_SNR_Review,IXPE_XRP_Review,dimarco2025b,Ursini2024_review}. 
X-ray emission from accreting compact objects exhibits strong variability on short timescales \citep{2006csxs.book...39V}. 
As a result, time-averaged X-ray polarimetry might risk smearing dynamic processes. 
Therefore, time dependent analysis is particularly needed for polarimetric studies of accreting compact objects \citep{bobrikova2024a,Mrk501_PA,xtej1701_pa,CirX_1_PA,2025A&A...693A.241Z} as well as jet-modinated emission in blazar \citep{DiGesu_Mrk421}. 

Recent observations with the Imaging X-ray Polarimetry Explorer \citep[IXPE;][]{Soffitta2021, weisskopf2022} have revealed fast variations of X-ray polarization in accreting compact objects, suggesting that the geometry and/or optical depth of the emission region (the Comptonization region specifically) is dynamically evolving.
For instance, \citet{bobrikova2024a} reported that GX 13+1 exhibits a significantly variable polarization degree (PD) ranging from 2\%-5\%, with the polarization angle (PA) rotating linearly by approximately $70^{\circ}$ over a course of two days. 
\citet{DiMarco2025} also reported a PA swing of $70^{\circ}$ between the dip and off-dip states.
Similarly, in Cir X-1, \citet{CirX_1_PA} observed a PA shift of \(49^{\circ} \pm 8^{\circ}\) along with the variation of the hardness ratio.
In another intriguing case, \citet{xtej1701_pa} and \citet{dimarco2025b} detected a rapid PA change during the normal branch of XTE J1701$-$462, whose polarization was otherwise undetectable in the time-averaged analysis due to cancellation of orthogonal components.  
The PA rotation in XTE J1701$-$462 is associated with variation in the reflection spectral component, suggestive of a rapid change in the Comptonization geometry between the boundary layer and the spreading layer. 
Alternatively, scattering in an extended accretion disk corona or disk wind may also contribute to the observed PA rotation \citep{DiMarco2025, Nitindala2025}.

These variations were observed over timescales of hours to days, to ensure that, as a common understanding, there are sufficient data counts in each time bin for a detection above the minimum detectable polarization (MDP). 
While for X-ray binaries, strong variability may occur on shorter timescales, in which case the number of source counts is insufficient to meet the above requirement. 
Therefore, it is worth exploring whether one can reliably and meaningfully determine the polarization in the case of low counts.
However, X-ray polarimetry is considered to be a photon starving technique, i.e., it requires a huge number of photons to establish a significant measurement of the PD. 
Conventionally, the MDP at 99\% confidence level is often used as a figure of merit to denote the sensitivity \citep{weisskopf2010}, and can be expressed in the case of negligible background as 
\begin{equation}
\text{MDP} \approx \frac{4.29}{\mu \sqrt{N}} \; ,
\end{equation}
where $\mu$ is the modulation factor and $N$ is the number of source photons. 
Given $\mu \approx 0.5$, a total number of $\sim$$10^6$ photons is needed to reach an MDP of 0.01. 
This is challenging for observations not only with current X-ray polarimeters like IXPE, but also with future larger telescopes like the enhanced X-ray Timing and Polarimetry \citep[eXTP;][]{zhang2018}.
Notably, for the investigation of fast variability via time-resolved analysis, this crisis cannot be alleviated by increasing the exposure time. 

In this paper, we argue that an accurate and useful constraint on the PA can be obtained in the case of low counts, i.e., a detection below MDP. 
In such a low statistics regime, due to the positive-definite nature of polarimetry, one must use a method that produces unbiased results, such as the Bayesian approach \citep{vaillancourt2006, quinn2012, maier2014, mikhalev2018}, which has been employed in the analysis of the data obtained with the small pathfinder PolarLight \citep{feng2020, long2021, long2022, long2023}. 
We note that this is different from unbinned analysis \citep{marshall2021, marshall2021a, marshall2024, 2023MNRAS.519.5902G}, which is not affected by the selection effect concerning bin size.
However, the unbinned analysis does not assist in identifying variation patterns in the result and a predefined model is always needed. 
For example, a linear rotation model was adopted in \citet{DiGesu_Mrk421} to test the PA variation in Mrk~421.

The paper is organized as following.
In Section~\ref{sec:meth}, we give a brief review of the Bayesian approach in X-ray polarimetry, particularly about how to constrain the PA\footnote{A PYTHON worksheet is available at 
\url{https://doi.org/10.5281/zenodo.15621003}}, in which example codes are provided for generating some of the results in this work. 
In Sections~\ref{sec:gx131}, \ref{sec:j1701}, and \ref{sec:scox1}, we apply the technique to three cases (GX 13+1, XTE J1701$-$462, and Sco X-1) as examples to illustrate how PA variations can be revealed. The results are discussed in Section~\ref{sec:disc}.

\section{Method}\label{sec:meth}

In this section, we briefly review the approach of polarimetric analysis based on the Stokes parameters \citep{kislat2015} and Bayesian inference \citep{quinn2012, maier2014, mikhalev2018}. 
More details can be found in the original papers.  
Assuming $\psi_i$ is the measured emission angle of the $i$th photoelectric event, the normalized Stokes parameters are defined as
\begin{align}
q_\text{m} = \frac{1}{N}\sum_{i=1}^{N}\cos2\psi_i \; , \\
u_\text{m} = \frac{1}{N}\sum_{i=1}^{N}\sin2\psi_i \; ,
\end{align}
where $N$ is the total number of events.
The measured PD $p_\text{m}$ and PA $\Psi_\text{m}$ can then be calculated as
\begin{align}
\label{equ:pm}p_\text{m} = \frac{2}{\mu}\sqrt{q_\text{m}^2 + u_\text{m}^2} \; , \\
\label{equ:psim}\Psi_\text{m} = \frac{1}{2}\arctan\frac{u_\text{m}}{q_\text{m}} \; ,
\end{align}
where $\mu$ is the mean modulation factor weighted by the measured source spectrum\footnote{In IXPE, the Stokes parameters $q_i$ and $u_i$ include a factor of 2 and thus the factor 2 in Equation~(\ref{equ:pm}) is not needed.}. 
The Stokes parameters $q$ and $u$ can be treated as Gaussian based on the central limit theorem when their 2D distribution is not truncated by the limit $q^2 + u^2 \le 1$. 
To ensure that, one usually requires $N$ is sufficiently large and/or PD is sufficiently small.
However, $p_{\rm m}$ and $\Psi_{\rm m}$ are no longer Gaussian due to nonlinear operations, and $p_{\rm m}$ is a biased estimate of PD.

\begin{figure}[t]
\centering
\includegraphics[width=\linewidth]{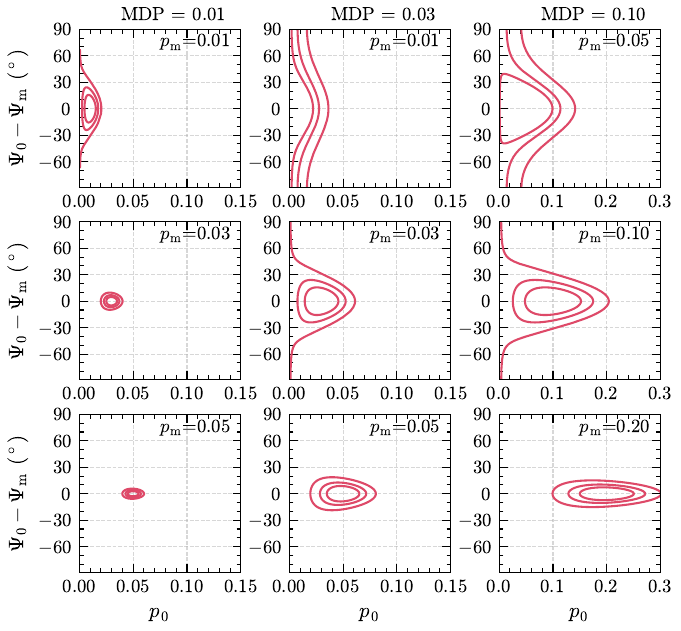}
\caption{Examples of bivariate posterior distributions of $(p_0, \Psi_0 - \Psi_\text{m})$ at different $p_\text{m}$ and MDPs. The contours represent the 68\%, 90\%, and 99\% confidence levels. A modulation factor $\mu = 0.3$ is assumed.}
\label{fig:posterior_2d}
\end{figure}

The Bayesian approach can provide an unbiased estimate of the intrinsic PD (the PA measurement is naturally unbiased). 
Here we use the subscript `m' to denote the measured quantities and `0' to denote the intrinsic values.
The posterior distribution $\rho(p_\text{0},\Psi_\text{0}|p_\text{m},\Psi_\text{m})$ can be computed with the likelihood $\rho(p_\text{m},\Psi_\text{m}|p_\text{0},\Psi_\text{0})$ and the prior distribution $\rho(p_\text{0},\Psi_\text{0})$ as
\begin{equation}
\rho(p_\text{0}, \Psi_\text{0} \mid p_\text{m}, \Psi_\text{m}) \propto 
\rho(p_\text{m}, \Psi_\text{m} \mid p_\text{0}, \Psi_\text{0})
\rho(p_\text{0}, \Psi_\text{0}) \; .
\end{equation}
The likelihood function describes the azimuthal distribution of the measured photoelectron emission angle and has a sinusoidal form. 
The prior distribution can be treated as non-informative and typically set to be uniformly distributed over the $p_0-\Psi_0$ plane, such that $\rho_\text{0}(p_\text{0},\Psi_\text{0}) = \text{const}$ \citep{quinn2012}. 
In most observations, two conditions are easily satisfied, $\mu^2p_\text{0}^2 \ll 1$ and $\sqrt{1/N} \ll 1$, allowing $q_\text{m}$ and $u_\text{m}$ to be treated as uncorrelated Gaussians. 
Then one has 
\begin{equation}
\begin{split}
&\rho(p_\text{m}, \Psi_\text{m} \mid p_\text{0}, \Psi_\text{0}) = \\ 
&\frac{p_\text{m}}{\pi \sigma^2} 
\exp\left\{ -\frac{p_\text{0}^2 + p_\text{m}^2 - 2 p_\text{m} p_\text{0} \cos \left[2(\Psi_\text{0} - \Psi_\text{m}) \right]}{2 \sigma^2} \right\},
\end{split}
\end{equation}
where $\sigma \approx \frac{1}{\mu} \sqrt{\frac{2}{N}}$ is the uncertainty of the measurement. 
Figure~\ref{fig:posterior_2d} shows examples of numerically computed bivariate posterior distributions of $(p_0, \Psi_0 - \Psi_\text{m})$ given different $p_\text{m}$ and MDPs. 

By integrating the 2D distribution along $\Psi_\text{0}$ or $p_\text{0}$, respectively, one obtains the marginalized posterior distribution, $\rho(p_\text{0}|p_\text{m})$ or $\rho(\Psi_\text{0}^\prime|p_\text{m})$, where $\Psi_\text{0}^\prime=\Psi_\text{0}-\Psi_\text{m}$. 
Point and interval estimations can be performed on the marginalized distributions numerically. 
The peak of the probability density function is commonly adopted as the point estimator of $p_\text{0}$, while the point estimator of $\Psi_\text{0}^\prime$ is always 0 since it is unbiased. 
The interval of the highest posterior density is adopted as the credibility interval.
We note that the estimate of $p_0$ is accurate enough (within 0.1\% and 1\%, respectively, for the point and interval estimates) if $N > 1000$, in comparison with the likelihood function that does not assume $q$ and $u$ are uncorrelated.

\begin{figure}[t]
\centering
\includegraphics[width=0.9\linewidth]{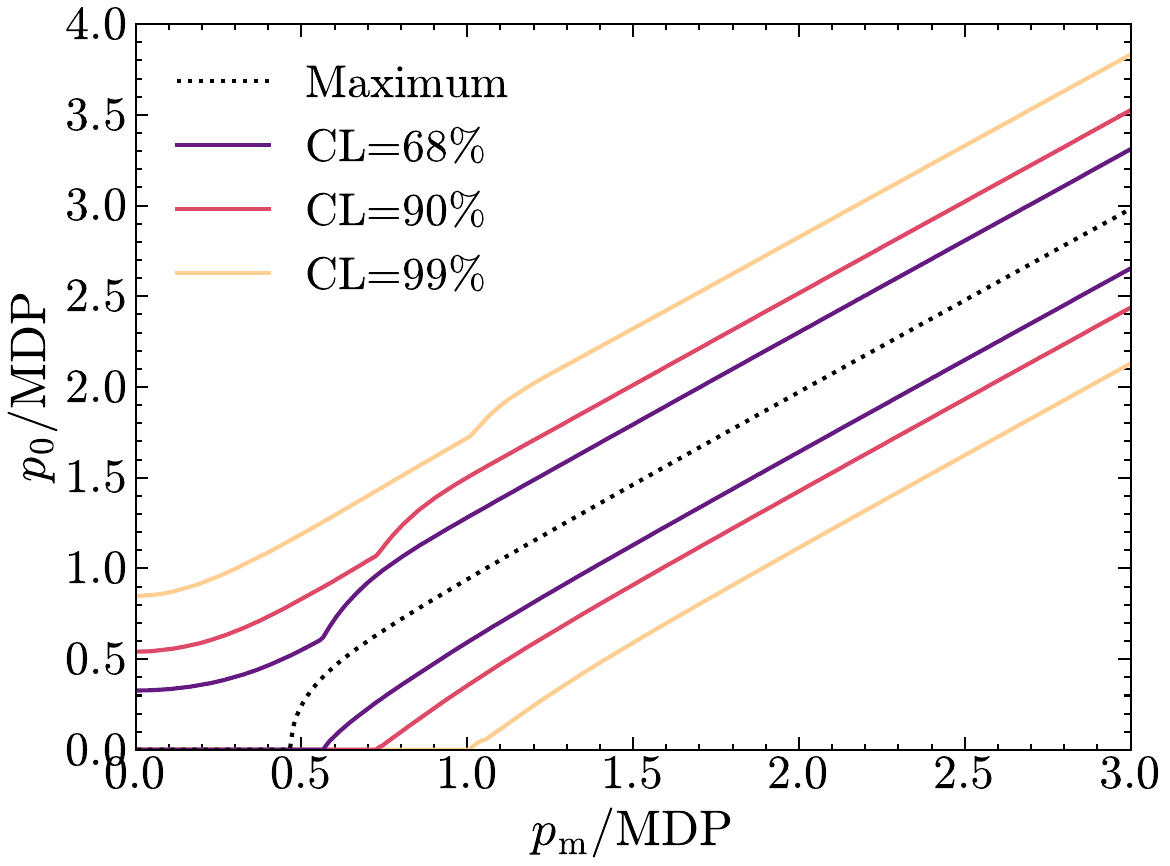}
\includegraphics[width=0.9\linewidth]{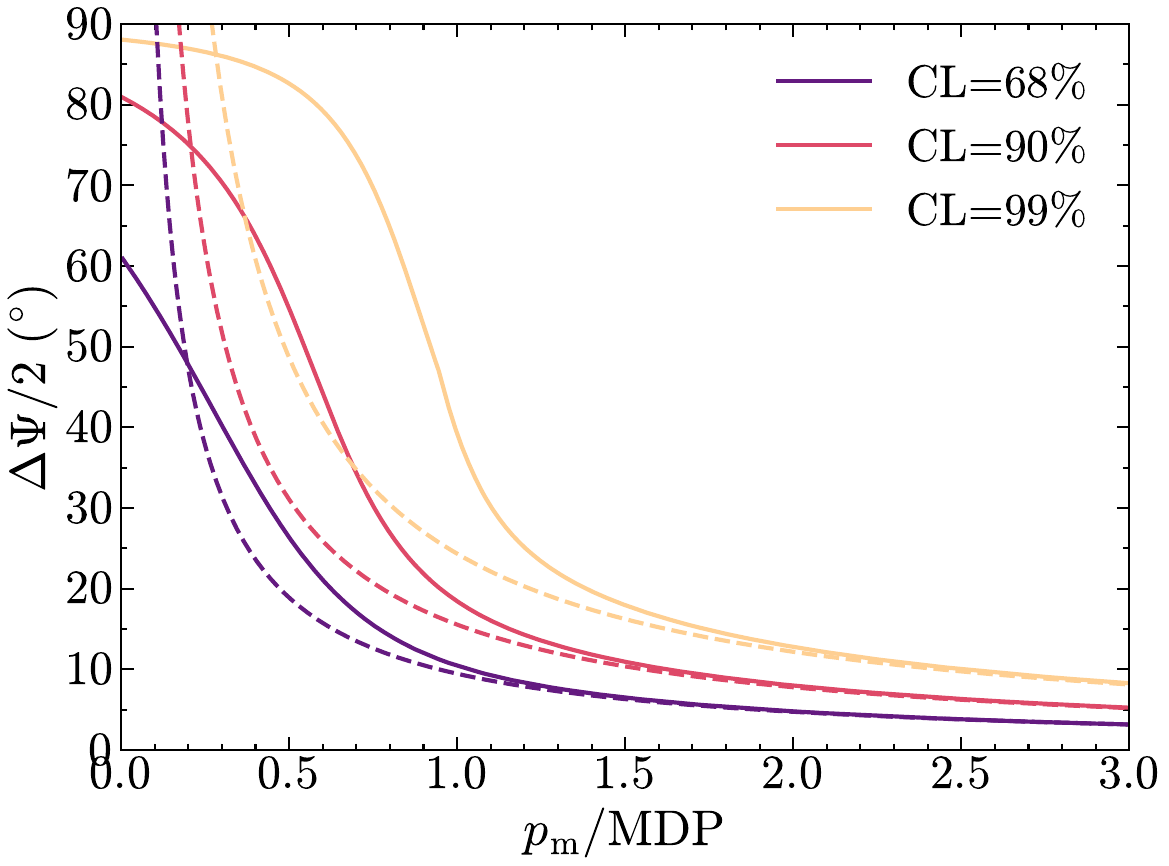}
\caption{Credibility intervals of $p_\text{0}$ and $\Psi_\text{0}$ at different confidence levels \citep[similar to Figs.~11 \& 12 in][]{maier2014}. 
The dashed lines in the bottom panel represent the corresponding PA uncertainties assuming Gaussianization in the high statistics regime \citep{kislat2015}. 
}
\label{fig:ci}
\end{figure}

\begin{figure}[t]
\centering
\includegraphics[width=0.9\linewidth]{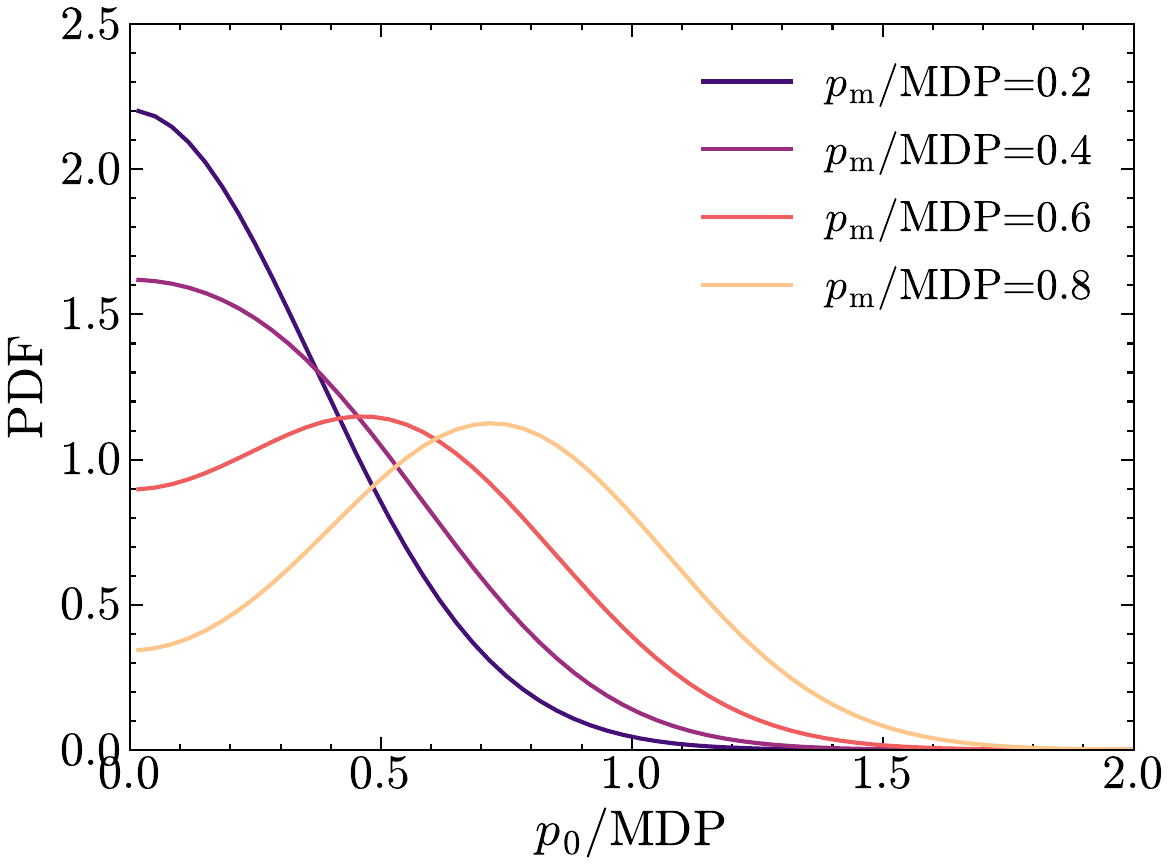}
\includegraphics[width=0.9\linewidth]{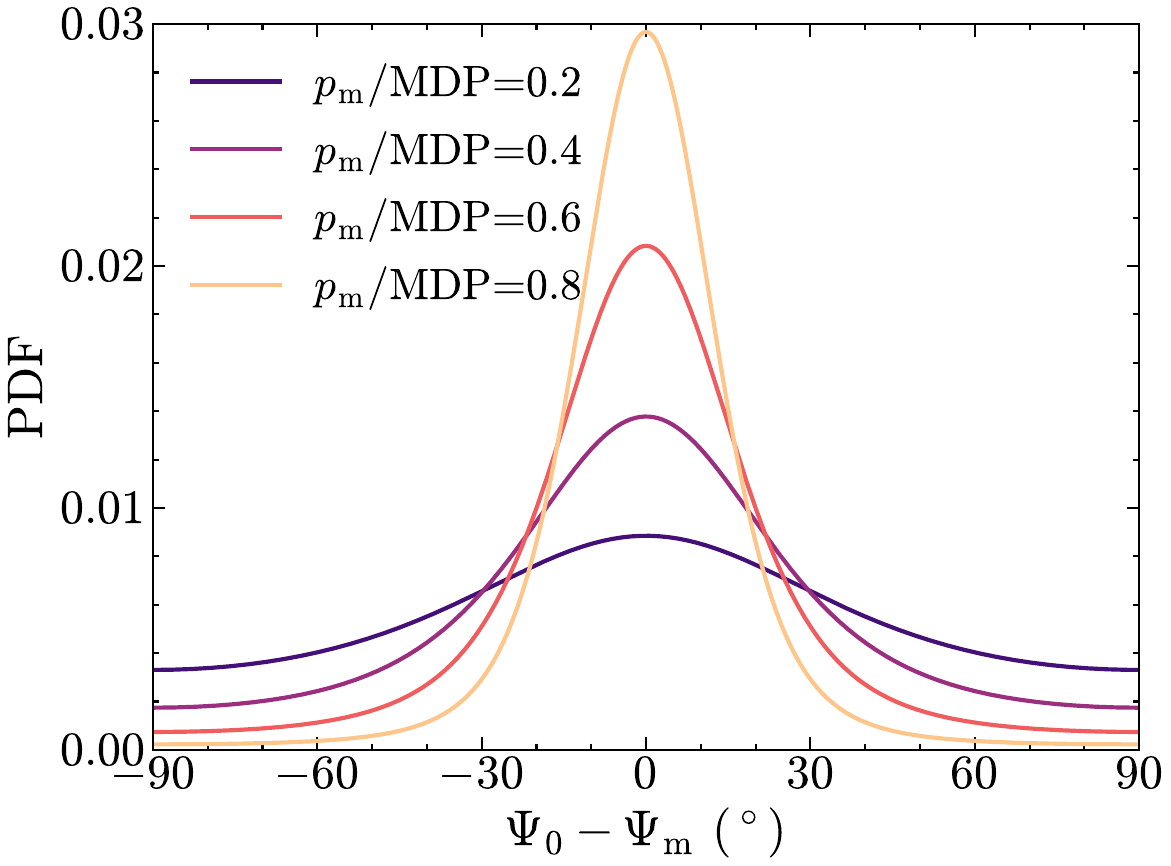}
\caption{Posterior distributions of $p_0/\text{MDP}$ \citep[similar to Fig.~9 in][]{maier2014} and $\Psi_0$ at different low-statistical levels, with $p_\text{m} / \text{MDP} =$ 0.2, 0.4, 0.6, and 0.8, respectively.}
\label{fig:pf0_lowsig}
\end{figure}

\citet{maier2014} pointed out that a neat form can be obtained if one uses the ratio of the polarization fraction to the measurement uncertainty, $\mathcal{P} \equiv p/\sigma$, as the parameter of interest, in which case $\rho(\mathcal{P}_0|\mathcal{P}_m)$ and $\rho(\Psi_\text{0}^\prime|\mathcal{P}_\text{m})$ are independent on $N$ and $\mu$. 
Here we use $p/\text{MDP}$ instead of $p/\sigma$, as MDP is proportional to $\sigma$ and is more commonly used in observations. 
Figure~\ref{fig:ci} shows the credibility intervals of $p_\text{0}$ and $\Psi_\text{0}$ at confidence levels of 68\%, 90\%, and 99\%, which are similar to Figs.~11 and 12 in \citet{maier2014}. 
Figure~\ref{fig:pf0_lowsig} shows the posterior distributions of PD \citep[similar to Fig.~9 in][]{maier2014} and PA, respectively, in cases of low counts, with $p_\text{m} / \text{MDP} < 1$. 

As one can see, in the case of low counts, the posterior distributions are no longer Gaussian and the traditional analysis tool like \texttt{ixpeobssim} \citep{baldini2022} cannot be used any more. 
Previous studies usually focus on the estimation of PD in these cases and have ignored the discussion on PA. 
Here we emphasize that, even if the PD is below MDP, a useful constraint on PA can still be obtained. 
In other words, the uncertainty of PA measurement is far below 180\arcdeg\ when $p_\text{m} / \text{MDP} =$ 0.5--1.0, allowing us to constrain the astrophysics to some extent, while one must bear in mind that the Bayesian approach must be used to correctly infer the PA uncertainty. 
Next, we apply the approach into the analysis of IXPE observations of three neutron star low-mass X-ray binaries (NS-LMXBs), GX 13+1, XTE J1701$-$462, and Sco X-1.

\section{GX~13+1}
\label{sec:gx131}

GX~13+1 is a persistent NS-LMXB located at a distance of 7~$\pm$~1~kpc \citep{bandyopadhyay1999}, exhibiting properties of both the atoll and Z sources \citep{schnerr2003, giridharan2024}. 
Four observations of GX~13+1 with IXPE have been conducted on 2023-10-17 (ObsID 02006801, Obs1), 2024-02-25 (ObsID 03001101, Obs2), 2024-04-20 (ObsID 03003401, Obs3), and 2025-09-08 (ObsID 04002901, Obs4) respectively.
The first three observations have been reported in the literature while the last one was performed during the review process of this paper.
A continuous PA rotation was previously reported in Obs1~\citep{bobrikova2024a}. Although no continuous rotation was observed in the subsequent two observations, a time dependence of the PA was also reported and discussed in Obs3~\citep{DiMarco2025}. Analyses resolved in hardness ratio and flux to investigate possible variations have also been performed. In those analyses, the data were binned with relatively large time intervals, e.g., at a timescale of 10--11~h for Obs1 by dividing the data into 5 segments. We try to examine if the conclusion remains with a finer time bin. 

We started the analysis with the level-2 data. 
Data reduction was performed with \texttt{ixpeobssim} 31.0.3. 
A circular region with a radius of 80$^{\prime\prime}$ was used for source extraction following \citet{bobrikova2024a}. 
Thanks to the brightness of the source, background subtraction was not performed as recommended by \citet{dimarco2023}. 
Source events in the 2--8~keV energy range are selected using \texttt{xpselect}. 
Considering the data gaps due to Earth occultation, we grouped the data obtained in each IXPE orbit as a trade-off between time resolution and statistics, with a time bin size of roughly 1~h and a cadence of about 1.5~h.

\begin{figure}
\centering
\includegraphics[width=0.9\linewidth]{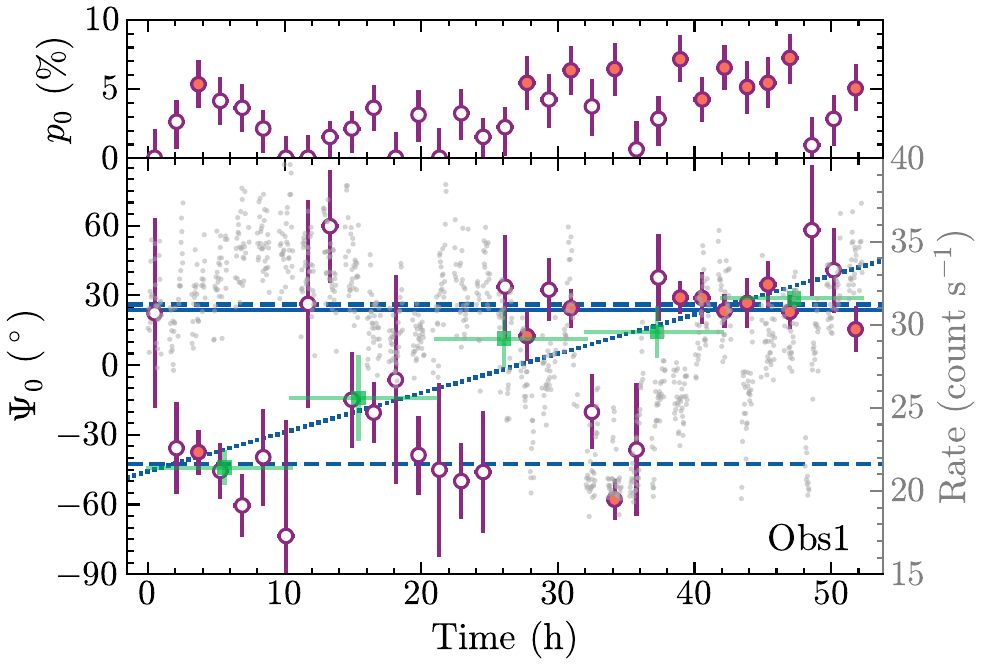}
\includegraphics[width=0.9\linewidth]{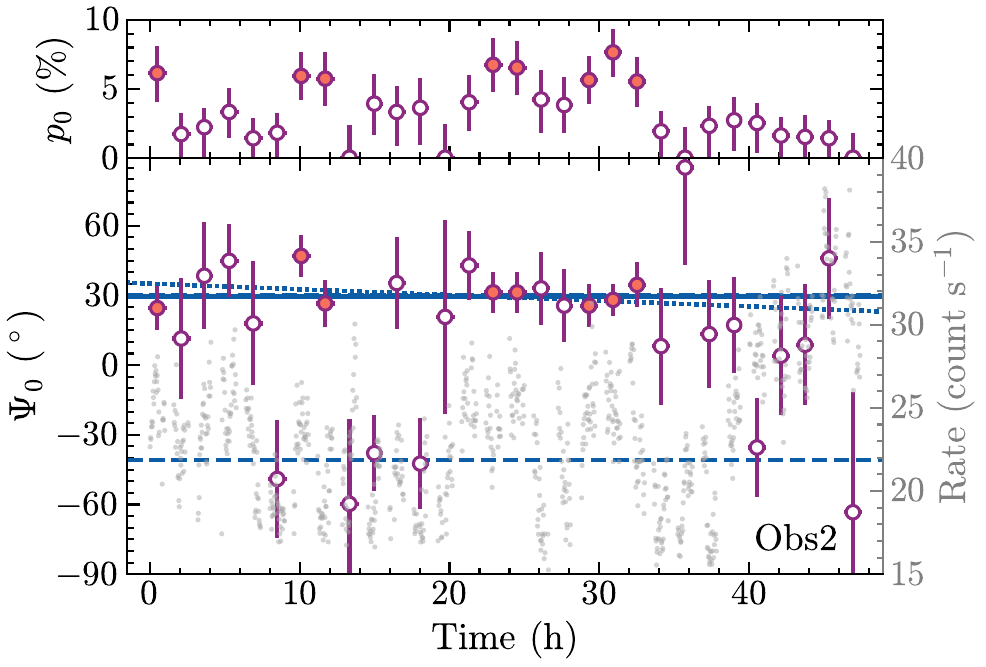}
\includegraphics[width=0.9\linewidth]{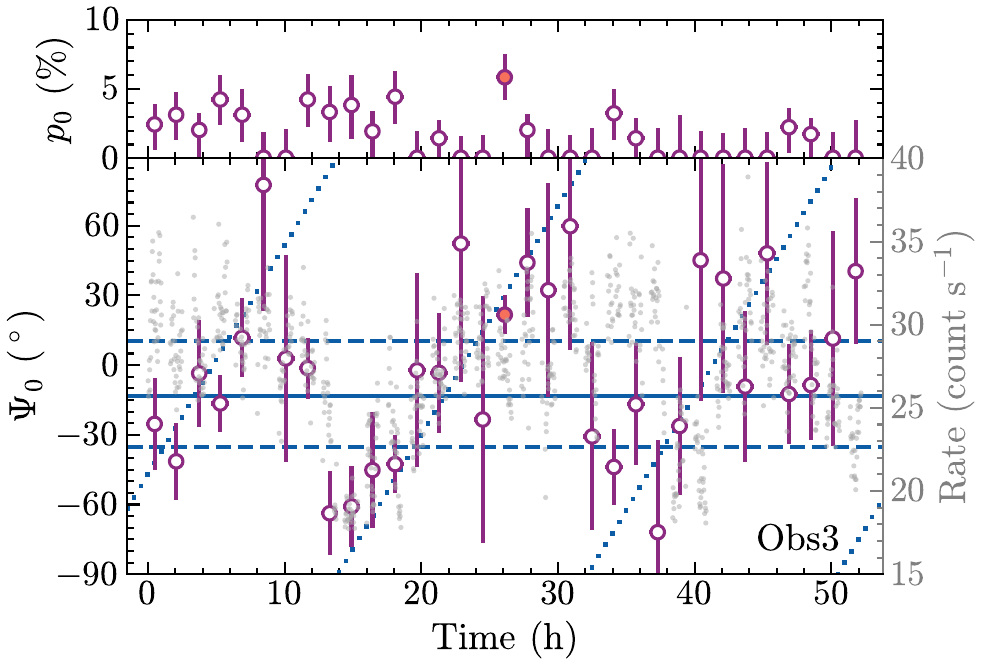}
\includegraphics[width=0.9\linewidth]{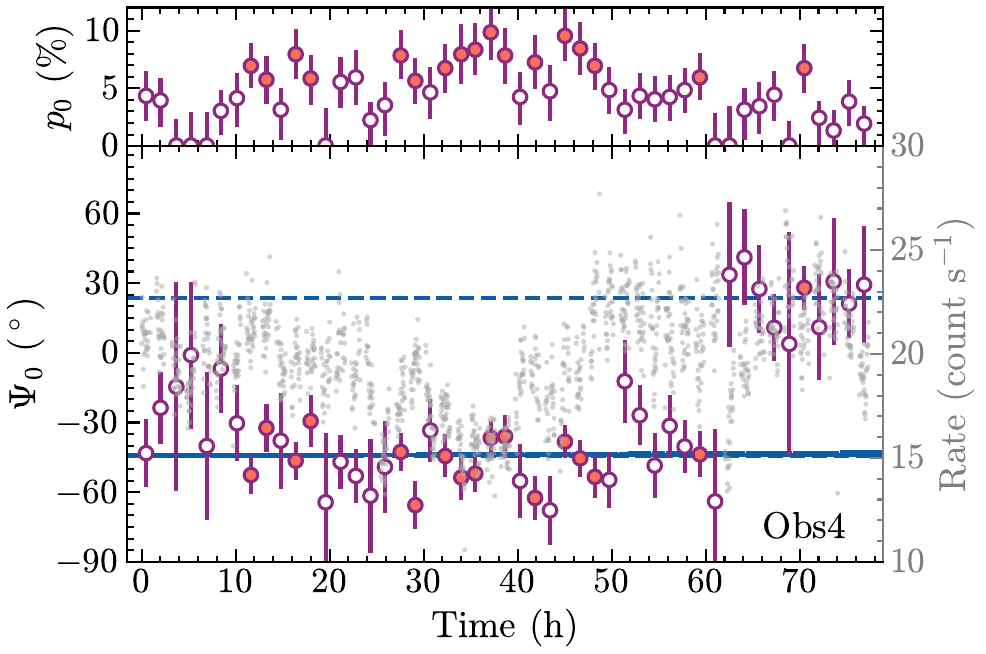}
\caption{Temporal variation of polarization properties and source count rate in the three IXPE observations of GX 13+1. Error bars indicate the 68\% credible intervals. Filled points indicate $p_{\rm m} / \mathrm{MDP} \geq 1$. In Obs1, the green data points are calculated using the binning scheme in \citet{bobrikova2024a}. The blue lines represent the three best-fit models: constant (solid), linear (dotted), and bimodal (dashed).}
\label{fig:pa0_gx131}
\end{figure}

First, we repeated the analysis described in \citet{bobrikova2024a} and obtained well consistent results.
Then, using the Bayesian approach introduced in Section~\ref{sec:meth}, we calculated the posterior distribution of $\Psi_\text{0}$ in each bin and inferred the PA and its 68\% error numerically.
The PA variations as a function of time in each observation are plotted in Figure~\ref{fig:pa0_gx131}.
There is no obvious trend of a linear rotation as reported by \citet{bobrikova2024a} in Obs1.
That is simply a result of the large time bin size that obscured the rapid variation.
We tested the results with three different models, a constant PA, a linearly rotating PA, and a bimodal PA, using the maximum likelihood estimation~(MLE), with the log-likelihoods defined as follows, 
\begin{align}
&\ln L_{\mathrm{con}}  = \sum_i \ln \mathrm{PDF}_i(c_0) \, , \\
&\ln L_{\mathrm{lin}} = \sum_i \ln \mathrm{PDF}_i(a + b\cdot t) \, , \\
&\ln L_{\mathrm{bim}} = \sum_i \ln \left[ \mathrm{PDF}_i(c_1)\cdot f_1 + \mathrm{PDF}_i(c_2)\cdot (1-f_1) \right] \, ,
\end{align}
where $c_0$, $c_1$, and $c_2$ are the PA, $b$ is the PA varying rate with time $t$, $a$ is the PA at time zero, $f_1$ is the probability of the PA appearing at $c_1$, and $\mathrm{PDF}_i$ is the posterior probability in the $i$th time bin.  
The best-fit results as well as the evaluation of models with the Akaike information criterion (AIC) and Bayesian information criterion (BIC) are listed in Table~\ref{tab:gx131}. 
We also computed the logarithmic marginal likelihood ($\ln Z$) assuming a uniform prior, presented in Table~1.
$\ln Z$ is insensitive to the sample size and allows us to compare the models using the logarithmic Bayes factor ($\ln {\rm BF} = \Delta \ln Z$).

\begin{deluxetable}{lllcccc}
\centering
\tabletypesize{\scriptsize}
\tablecaption{Best-fit parameters, logarithmic likelihood, AIC, BIC, and logarithmic marginal likelihood for the three models in the analysis of GX 13+1.}
\label{tab:gx131}
\tablehead{
\colhead{Obs} & \colhead{Model} & \colhead{Parameters} & \colhead{$\ln\mathcal{L}$} & \colhead{AIC} & \colhead{BIC} & \colhead{$\ln Z$}}
\startdata
 & const. & $c_0 = 23.7$ & $-$187.2 & 376.4 & 377.9 & $-$190.6\\
Obs1 & linear & $b = 1.69$, $a = -45.8$& $-$164.5 & 333.0 & 336.0 & $-$171.1\\
 & bimodal & $c_1 = -42.7$, $c_2 = 26.1$, $f_1 = 0.45$ & $-$154.4 & 314.7 & 319.2 & $-$161.6 \\ 
\hline
 & const. & $c_0 = 29.6$ & $-$137.4 & 276.8 & 278.2 & $-$140.8\\
Obs2 & linear & $b = -0.25$, $a = 35.1$ & $-$136.8 & 277.6 & 280.4 & $-$143.0\\
 & bimodal & $c_1 = -40.7$, $c_2 = 30.0$, $f_1 = 0.15$ & $-$133.4 & 272.9 & 277.1 & $-$139.9\\
\hline
 & const. & $c_0 = -13.3$ & $-$171.9 & 345.8 & 347.3 &$-$174.2\\
Obs3 & linear & $b = 9.84$, $a = -46.9$ & $-$168.9 & 341.8 & 344.8 &$-$173.5\\
 & bimodal & $c_1 = -35.3$, $c_2 = 10.4$, $f_1 = 0.53$ & $-$164.9 & 335.9 & 340.4 & $-$169.4\\
\hline
 & const. & $c_0 = -43.8$ & $-$232.8 & 467.7 & 469.6 & $-$236.6\\
Obs4 & linear & $b = 0.02$, $a = -44.5$ & $-$232.8 & 469.7 & 473.5 &$-$240.7\\
 & bimodal & $c_1 = -44.5$, $c_2 = 23.6$, $f_1 = 0.82$ & $-$219.8 & 445.7 & 451.3 & $-$227.2\\
\enddata
\tablecomments{
$c_{0,1,2}$ are the PAs in the unit of degree. $b$ is the slope in the unit of degree per hour and $a$ is the PA at time zero in degree. $f_1$ is the probability of PA appearing at the $c_1$.}
\end{deluxetable} 

Instead of a continuous or linear PA rotation, our results at a smaller time bin favor the scenario that the PA swings between two discrete angles, one around $-43\arcdeg$ and the other around $26\arcdeg$, by a difference of roughly $70\arcdeg$, in particular in Obs1.  
Such a bimodal model surpasses the linear model given the AIC or BIC tests, as well as the Bayes factor, $\ln {\rm BF} = 9.5$, indicative of ``very strong'' evidence \citep{kass1995}. 
For Obs2, the constant PA model cannot be significantly rejected compared with the bimodal model ($\ln {\rm BF} = 0.9$).
This is mainly because the PA appears at the other value only in a few observations. 
These are consistent with the results reported in \citet{bobrikova2024b}.
For Obs3, the constant model is clearly disfavored compared with the bimodal model ($\ln {\rm BF} = 4.8$).
However, one should be cautious if the PA distribution is exactly `bimodal', which is preferred because additional parameters help improve the fit, but the exact pattern of PA distribution cannot be well constrained with the current data. 
In Obs3, the linear model results in a significantly higher angle rotation rate ($b \approx 9.8^\circ / \text{h}$, see Table~\ref{tab:specfit}), causing multiple angle warps during the observation.
However, the linear model in Obs3 is not significantly preferred over the constant model given the small difference in AIC or BIC. 
We then fitted the $q$ and $u$ values, which are presumably Gaussian, with the linear model and obtained an improvement of $\Delta \chi^2 = 1.9$ over the constant model with one additional degree of freedom, suggesting that the linear model is not significantly better, consistent with the Bayesian analysis. 
We note that the $\chi^2$ analysis in the $q-u$ space is not applicable to the bimodal model as $f_1$ is a probability, lacking a one-to-one correspondence between data and model. 
For Obs4, only the DU1 and DU3 data are released for analysis at the moment. 
The bimodal model is preferred over the other two ($\ln {\rm BF} = 9.4$ or $13.5$). 
Remarkably, the best-fit PAs are well consistent with those inferred with Obs1.

\begin{figure}[tb]
\centering
\includegraphics[width=0.9\linewidth]{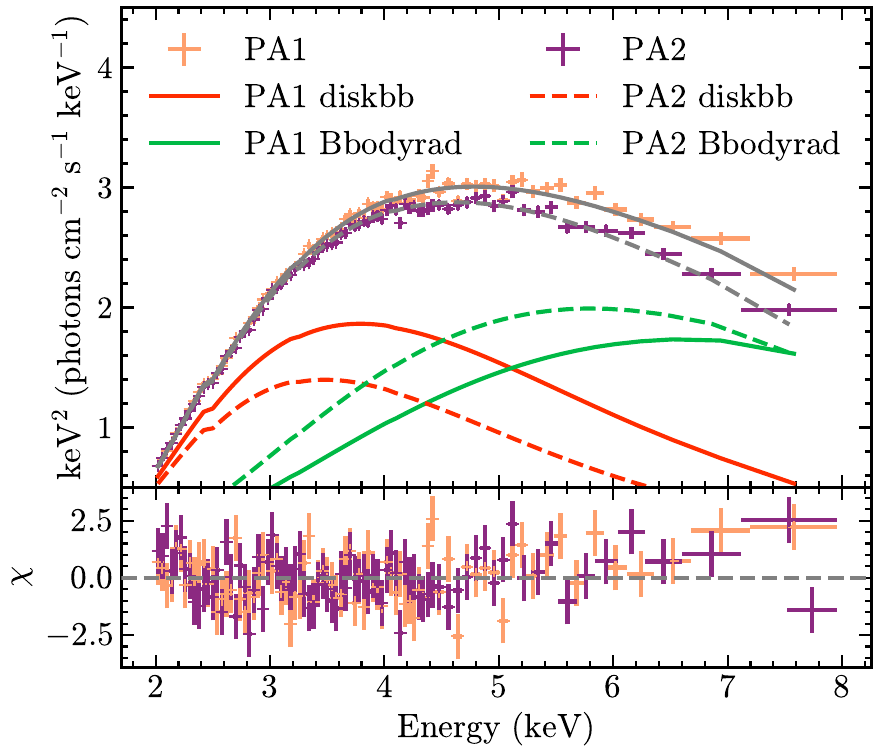}
\caption{Energy spectra of GX 13+1 in the two states with different PAs, where $\text{PA}1 \approx -43\arcdeg$ and $\text{PA}2 \approx 26\arcdeg$. Only the IXPE DU1 data are plotted for clarity.}
\label{fig:specfit}
\end{figure}

To see if there is any related spectral variation along with the PA variation, we extracted the energy spectra in time intervals associated with each PA in Obs1. 
We constructed the good time intervals (GTIs) based on the PA obtained from the analysis described above, and extracted the events using the \texttt{xselect} tool within the GTIs. 
Energy spectra are then generated using the \texttt{PHA1} algorithm implemented in \texttt{xpbin}, employing the unweighted method and response files from version \texttt{20230702}, grouped with a minimum of 25 counts per bin.
The spectra are fitted with the \texttt{Tbabs(diskbb+bbodyrad)} model in \texttt{XSPEC}, a phenomenological model in combination with a blackbody and a disk blackbody, subject to interstellar absorption. 
The two spectra are shown in Figure~\ref{fig:specfit}, and the best-fit parameters are listed in Table~\ref{tab:specfit}.
The best-fit parameters are consistent with each other considering the uncertainties.
Furthermore, to investigate if there is any correlation between the PA and flux, we plotted the source count rate in Figure~\ref{fig:pa0_gx131}. 
No apparent correlations can be revealed in Obs1 and Obs2.
However, the PA in Obs3 seems to be scaled with the source count rate \citep[see also Fig.~5 in][]{DiMarco2025}.

\begin{deluxetable}{llcc} 
\tabletypesize{\footnotesize}
\tablewidth{\columnwidth}
\tablecaption{Best-fit parameters of spectral fitting for GX~13+1.}
\label{tab:specfit}
\tablehead{
\colhead{Model}  & \colhead{Parameter} & \colhead{${\rm PA} \approx -43 \arcdeg$} & \colhead{${\rm PA} \approx 26 \arcdeg$}
}
\startdata
      Tbabs & $N_{\rm H}$ ($10^{22} \ \rm cm^{-2}$) & $5.1^{+0.1}_{-0.2}$ & $5.2^{+0.3}_{-0.2}$ \\
      Diskbb   & $T_{\rm in}$ (keV) & $1.2^{+0.3}_{-0.2}$ & $1.0^{+0.3}_{-0.1}$ \\
         & Norm & $233^{+240}_{-115}$ & $382^{+311}_{-223}$ \\
      Bbodyrad   & $kT$ (keV) & $1.6^{+0.8}_{-0.2}$ & $1.4^{+0.3}_{-0.1}$ \\ 
         & Norm & $51^{+64}_{-45}$ & $116^{+62}_{-89}$  \\ 
         \noalign{\smallskip}\hline\noalign{\smallskip}
        Cross-cal   & $C_{\rm du1}$  & 1$^\ast$ & 1$^\ast$ \\
          & $C_{\rm du2}$  & $1.017\pm0.003$ & $1.016\pm0.003$ \\
           & $C_{\rm du3}$  & $0.999\pm0.003$ & $0.999\pm0.003$\\
         \noalign{\smallskip}\hline\noalign{\smallskip}
         & $\chi^2 / {\rm d.o.f}$ & 432.66/437 & 486.56/437 \\ 
\enddata
\tablenotetext{^\ast}{Parameters fixed in the fit.}
\tablecomments{Uncertainties are quoted at the 90\% confidence level.}
\end{deluxetable}

\section{XTE J1701$-$462}\label{sec:j1701}

XTE J1701$-$462 is a transient NS-LMXB at a distance of $8.8 \pm 1.3$~kpc \citep{lin2009}.  
The source was observed twice with IXPE in 2022 September during an outburst, in the horizontal branch (HB) and normal branch (NB), respectively.

During the HB observation (ObsID: 01250601), a relatively high PD of approximately 4.6\% was detected, with a PA of about $-$38$^{\circ}$ \citep{cocchi2023}. 
For the NB observation (ObsID: 01250701), initial studies reported a nondetection of polarization \citep{cocchi2023,yu2025}. 
However, subsequent studies by dividing the data into three \citep{xtej1701_pa} or five \citep{dimarco2025b} time bins identified a rapid PA variation within the observation that has lead to depolarization and nondetection in the time-averaged data.

\begin{figure}[t]
\centering
\includegraphics[width=\linewidth]{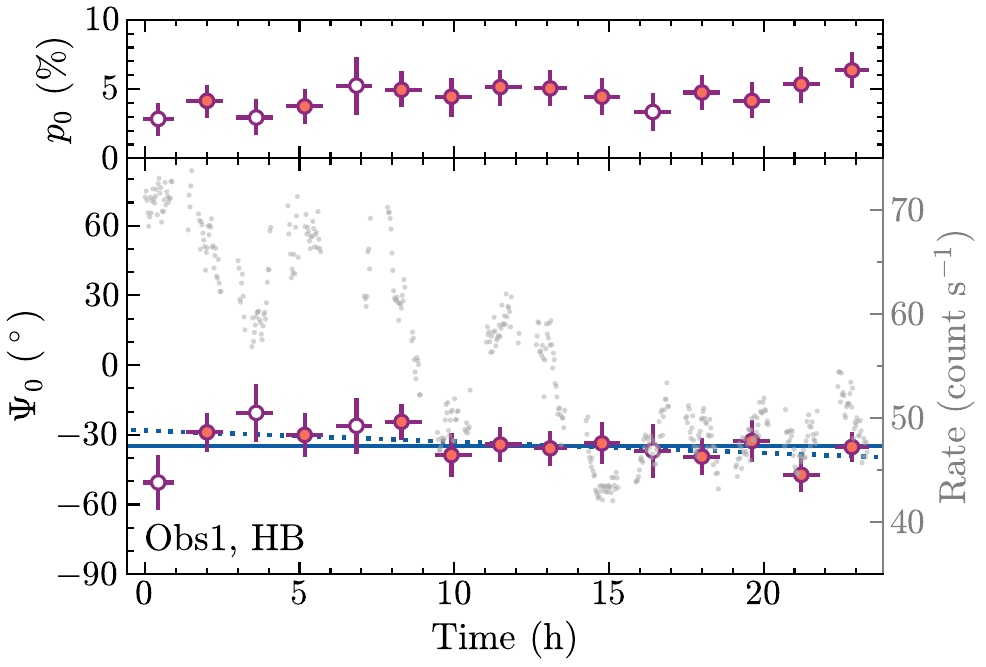}
\includegraphics[width=\linewidth]{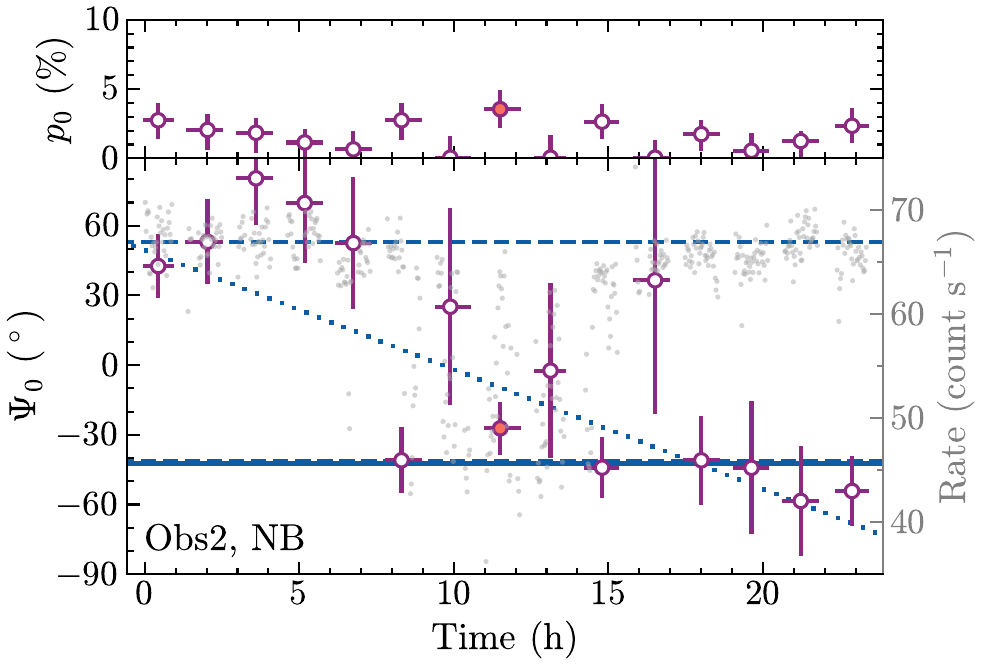}
\caption{Temporal variation of polarization properties and source count rate in the two IXPE observations of XTE J1701$-$462, in the same format as in Figure~\ref{fig:pa0_gx131}.
Error bars indicate the 68\% credible intervals. Filled points indicate $p_{\rm m} / \mathrm{MDP} \geq 1$.}
\label{fig:j1701_pa_lc}
\end{figure}

We reanalyzed the data in the same manner as for GX 13+1. A circular region with a radius of 90\arcsec\ was used for source extraction to be consistent with the approaches used in \citet{xtej1701_pa}. 
Here we show PA variation at a time step of the IXPE orbit (15 time bins), with results shown in Figure~\ref{fig:j1701_pa_lc}.
Our results are consistent with those reported in the literature, confirming the presence of PA variation in the NB.
Furthermore, our results at a finer time resolution reveal that the PA variation may have occurred on a timescale as short as 1.5 hours (a single IXPE orbit).
We fitted the PA variation with the same three models as for GX 13+1. 
The best-fit results are listed in Table~\ref{tab:j1701}. 
For Obs1, the linear and bimodal models are not preferred over the constant model given the AIC, BIC, and \(\ln(Z)\). 
For Obs2, the constant model can be rejected ($\ln {\rm BF} = 4.9$ or $6.2$).
The AIC and $\ln (Z)$ suggest that the bimodal model is the best one, but quite comparable with the linear model, while the BIC indicates both the linear and bimodal models are equally preferred. 
Therefore, we suggest that the current data cannot significantly distinguish between the two models. 

A detailed broadband spectral analysis of XTE J1701$-$462 using IXPE and NuSTAR data has already been performed and presented in \citet{xtej1701_pa}. The analysis revealed enhanced disk and reflection emission in the early epoch of Obs2.

\begin{deluxetable}{lllcccc}
\centering
\tabletypesize{\scriptsize}
\tablecaption{Best-fit parameters, logarithmic likelihood, AIC, BIC, and logarithmic marginal likelihood for the three models in the analysis of XTE J1701$-$462.}
\label{tab:j1701}
\tablehead{
\colhead{Obs} & \colhead{Model} & \colhead{Parameters} & \colhead{$\ln\mathcal{L}$} & \colhead{AIC} & \colhead{BIC} & \colhead{$\ln Z$}}\startdata
& const. & $c_0 = -34.7$ & $-$51.4 & 104.8 & 105.5 & $-$55.0\\
Obs1 & linear & $b = -0.48$, $a = -28.1$ & $-$50.1 & 104.2 & 105.6 & $-$57.5\\
 & bimodal & $c_1 = c_2 = -34.7$, $f_1 =arb.$ & $-$51.4 & 108.8 & 110.9 & $-$56.3\\
\hline
 & const. & $c_0 = -42.4$ & $-$80.8 & 163.6 & 164.4 & $-$83.4\\
Obs2 & linear & $b = -5.14, a = 49.5$ & $-$73.7 & 151.4 & 152.8 & $-$78.5\\
 & bimodal & $c_1 = 53.0, c_2 = -41.4, f_1 = 0.40$ & $-$72.4 & 150.7 & 152.9 & $-$77.2\\
\hline
\enddata
\end{deluxetable} 

\section{Sco X-1}
\label{sec:scox1}

Sco X-1 is the first accreting compact object with a significant X-ray polarization measurement \citep{long2022}. 
Both PolarLight and OSO-8 measured a time-averaged PA in line with the jet orientation on the sky plane (\citealt{long2022}; \citealt{long1979}). 
In particular, PolarLight observations suggest that the polarization is more significant when the source count rate is higher. 
However, an IXPE observation discovered a different PA with a short observation \citep{monaca2024, lamonaca2025}, triggering speculation if the source has experienced a PA variation like those observed in many other NS-LMXBs. 

\begin{figure}[tb]
\centering
\includegraphics[width=0.9\linewidth]{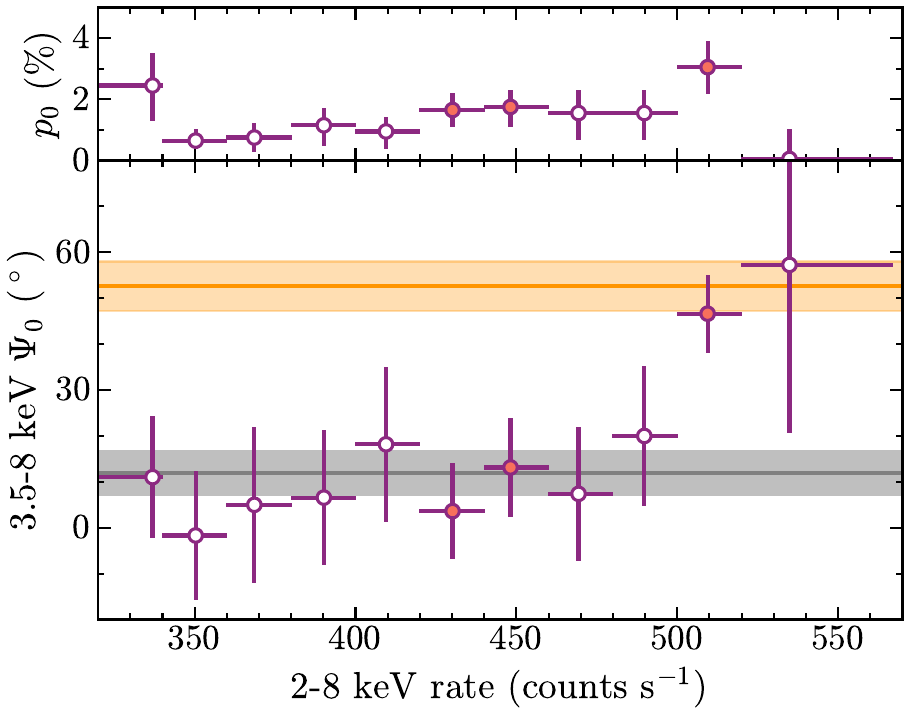}
\caption{PA variation as a function of source count rate for Sco X-1. The source flux is quoted in 2--8 keV while the polarization is calculated in 3.5--8 keV. The gray line and shaded region represent the time-averaged IXPE result \citep{monaca2024} and the yellow line and shaded region represent the PolarLight measurement \citep{long2022}. 
Error bars indicate the 68\% credible intervals. Filled points indicate $p_{\rm m} / \mathrm{MDP} \geq 1$.}
\label{fig:scox1_pa_rate}
\end{figure}

The data reduction follows the same procedure as for GX 13+1, and a circular region with a radius of 80$^{\prime\prime}$ was used for source extraction.
We divided the IXPE data into different flux bins, with results shown in Figure~\ref{fig:scox1_pa_rate}, to investigate if PA correlates with the source flux.  
When the IXPE count rate in the 2--8~keV band is below $\sim$500~\cts, the measured PA is consistent with the time-averaged value. 
When the rate exceeds 500~\cts, the PA shifts to a level consistent with the PolarLight measurement. 
We calculated the Bayes factor by comparing the model in which PA changes above 500~\cts, against the model with a constant PA. 
We performed the analysis in different energy bands, and obtained $\ln {\rm BF} = $ 0.57, 5.3 and 2.2, respectively, in the enery range of 3--8, 3.5--8, and 4--8~keV.
The results suggest that the evidence for variation is not significant at all in 3--8 keV, positive in 4--8 keV, and very strong in 3.5--8~keV \citep{kass1995}. 
This is a hint that the PA variation with flux may be energy dependent, and indicates that simultaneous spectral and polarimetric analysis is needed to resolved this problem.
However, we are not able to find an independent criterion to choose the energy band for polarimetric analysis. One should keep the caveat in mind.  

Also, we computed the $q$ and $u$ values and their variances for data below and above 500~\cts.
In the energy band of 3.5--8 keV, with the Hotelling's T-squared test, one gets $\chi^2 = 12.5$ with 2 degrees of freedom, corresponding to a $p$-value of 0.0019, compatible with the above conclusion.

\section{Discussion and conclusion}
\label{sec:disc}

In this paper, we demonstrated that a useful constraint on PA is still possible with X-ray polarimetry in the regime of low statistics ($p_{\rm m} \lesssim {\rm MDP}$), where the Bayesian approach must be used for an accurate estimate.
Standard analysis that assumes Gaussian distributions will lead to overestimation of PA in this regime, because the prior information of PA in 0\arcdeg--180\arcdeg\ is not considered.
In this case, the constraint on PA is in practice more useful than on PD because it has a smaller relative error --- the intrinsic PD in nonmagnetic systems is usually below 5\% while the dynamical range of PA is always 0--180\arcdeg.
Thus, a PD upper limit of 5\% or 10\% is not particularly effective at physical modeling, but a PA uncertainty of 10\arcdeg\ or even 30\arcdeg\ is quite helpful in constraining the physics.
Without Gaussianization, the maximum likelihood instead of $\chi^2$ analysis should be employed.
We applied the methodology to several NS-LMXBs for examples.

In the first and third observations of GX 13+1, a constant PA can be ruled out. 
Moreover, our results indicate that the PA variation in Obs1 is not in a mode of linear rotation, but swing between two values, also consistent with the PA distribution seen in Obs2 and Obs4.
A similar bimodal PA swing may have also occurred in XTE J1701$-$462 during Obs2, but at a PA difference close to 90\arcdeg. 
The nature of such a PA variation is unclear.
In the first scenario, there could be two emitting components, e.g., a boundary layer and a spreading layer, competing with each other along with the variation of the accretion rate.
Or, in the second scenario, there is only one dominant Comptonization component that is varying its geometry and/or optical depth in response to the change of accretion rate, leading to a PA variation. 
The bimodal variation mode, if confirmed with future observations, seems to favor the first scenario.
In this case, the associated spectral variation along with PA swing, may suggest that the emission from the spreading layer is harder.

Here we emphasize that the PA cannot be directly associated with geometry --- PA is not simply aligning with or perpendicular to the elongation of the Comptonization region --- the number of scattering matters as well. 
Assuming there is an elliptical Comptonization region, when the scattering mean free path is comparable to the geometric size, the average PA tends to be perpendicular to the long axis, as there is a higher chance that the last scattering occurs along the long axis. 
When the scattering mean free path is much smaller than the geometric size, the average PA tends to be in line with the long axis, similar to the case of stellar atmosphere as discussed in \citet{Chandrasekhar1960}. 
Any PA in between is possible when the number of scattering varies from a few (optical depth $\tau$ $\sim$ unity or a few) to a great number ($\tau \gg 1$). 
Sometimes, a simple optical depth cannot tell the number of scattering, as the number also depends on the seed and electron temperatures and the emergent energy.
Thus, a switch between the boundary layer and spreading layer does not necessarily lead to a PA swing of 90\arcdeg.

In both GX 13+1 (Obs3) and Sco X-1, a possible correlation is seen between the PA and source flux, suggesting that the accretion rate is the main driver of the variation of the Comptonization geometry and/or optical depth. 
This is consistent with our understanding of NS-LMXBs that an increasing accretion rate may lead to squeezing of the boundary layer and expanding of the spreading layer. 
However, it is unclear why and how the source experiences two different PA variation patterns, i.e., the bimodal pattern and the PA-flux correlated pattern. 
We also note that a similar phenomenon in GX 13+1 has been noticed by \citet{DiMarco2025}, that there is a PA swing of about 70\arcdeg\ between the dip and non-dip time intervals, corresponding to the bimodal PA swing ($\Delta {\rm PA} \approx 70\arcdeg$) in this work. 
They interpreted the PA swing associated with flux dipping as due to partial obscuration of the disk corona.
However, we point out that, as one can see in Obs1 (Figure~\ref{fig:pa0_gx131}), a PA of around $-40\arcdeg$ during flux dips is also seen at non-dipping intervals. 
Therefore, more observations are needed to firmly determine if the PA is correlated with flux or dips.

An extreme case of data binning is the single-event (unbinned) analysis \citep[see a similar case in][]{2023MNRAS.519.5902G}.
One can compute the $\Psi_0$ distribution for each event and perform a maximum likelihood analysis. 
In that case, a specific model is needed a priori. 
The single-event analysis is not able to predict the model, while the binned analysis can, as demonstrated by the bimodal model example in this case. 

\begin{acknowledgments}
We thank the referees for useful comments.
We acknowledge funding support from the National Natural Science Foundation of China (NSFC) under grant Nos. 12025301 and 12122306, and the Strategic Priority Research Program of the Chinese Academy of Sciences. 
\end{acknowledgments}

\bibliographystyle{aasjournal}

\end{document}